# Mesophase formation in a system of top-shaped hard molecules:

# Density functional theory and Monte Carlo simulation


**Daniel de las Heras[a), Szabolcs Varga[b) and Franz J. Vesely[c)**

[a) *Centro de Física Teórica e Computacional da Universidade de Lisboa, Av. Profesor Gama Pinto 2, P-1749-016, Lisbon, Portugal*

[b) *Department of Physics and Mechatronics, University of Pannonia, H-8200 Veszprém, PO Box 158, Hungary*

[c) *Computational Physics Group, Faculty of Physics, University of Vienna, Boltzmanngasse 5,  A-1090 Wien, Austria*


Number of pages: 30 (including figure captions, 9 figures and 2 tables)


[a) e-mail: delasheras.daniel@gmail.com

[b) e-mail: vargasz@almos.vein.hu

[c) e-mail: franz.vesely@univie.ac.at




## Abstract


We present the phase diagram of a system of mesogenic top-shaped molecules based on the Parsons-Lee density functional theory and Monte Carlo simulation. The molecules are modeled as a hard spherocylinder with a hard sphere embedded in its center. It is found that the central spherical unit destabilizes the nematic with respect to the isotropic phase, while increasing the length of the cylinder has the opposite effect. Also, the central hard sphere has a strong destabilizing effect on the smectic A phase, due the inefficient packing of the molecules into layers. For large hard sphere units the smectic A phase is completely replaced by a smectic C structure. The columnar phase is first stabilized with increasing diameter of the central unit, but for very large hard sphere units it becomes less stable again. The density functional results are in good agreement with the simulations.




# 1.     Introduction

Understanding the relation between molecular structure and macroscopic properties is a very important issue, because many practical applications require precise tailoring of the phase behavior. Nowadays very complex mesogenic molecules are available to study the link between the observed mesophases and the molecular structure[1]. For example, mesogenic (rod, plate) and non-mesogenic (sphere, flexible polymer) building blocks can be attached together to study the role of different parameters such as the shape anisotropy, the polarizability and the flexibility in the stabilization of liquid crystalline phases. The only problem with these complex molecular formations (for example: rod-coil, coil-rod-coil molecules) is the presence of very complex interactions between the different building blocks, which do not allow us to identify the separate roles of different molecular interactions in the stabilization of the mesophases. One way to overcome this problem is to construct model systems that can be studied by theory and simulation.

Rod-shaped molecules such as hard spherocylinders have been shown to self-assemble into a rich variety of different liquid crystalline structures through the anisotropic repulsive interactions. The mesophases include nematic, smectic A and solid phases[2]. The shape of many mesogenic particles such as some metallomesogenic molecules[3] and fullerene containing calamitic liquid[4] crystals are closer to top-shaped bodies. The use of hard body models for these molecules helps us to understand the role played by the steric repulsive forces in the stability of different mesophases.

The first theoretical work on top-shaped molecules, which was devoted to studying the effect of varying the central sphere diameter on the nematic and smectic A phases, was done by Cinacchi[5]. In this study the central core is modeled as a soft sphere and the Parsons-Lee[6] extension of the Onsager theory[7] is applied. Interestingly, it is found that making the central core bulkier can enhance the stability of the nematic and smectic A phases. Luckhurst[8] studied a thermotropic system of spherocylinders with an embedded sphere using Monte Carlo simulation. Instead of stabilization of the smectic A phase, the system stays in the nematic phase, and a strong tendency for forming a tilted smectic phase is observed. In the study of Kim et al.[9] the phase behavior of 11-site Lennard-Jones



(LJ) linear molecules is examined using molecular dynamic simulation. Instead of one central LJ sphere, three LJ sites are chosen in the centre of the rod-like molecule to make it more bulky. It is found that interdigitated smectic order replaces the conventional smectic A phase. Very recently fullerene containing triblock mesogens (a central fullerene with two mesogenic groups) has been examined by Orlandi et al[10], where the stabilization of the smectic order is detected in a layered structure such that the fullerene spheres and mesogenic rods are microsegregated. In several theoretical studies an attractive square-well site is placed in the centre of the mesogenic unit to see the effect of the square-well radius on the stability of the isotropic-nematic and nematic-smectic phase transitions. Nematic-nematic phase transition[11], stabilization of the smectic ordering[12] and nematic reentrant behavior[13] are observed. The model has been applied successfully for p-azoxianisol[14]. What the above studies have in common is that the systems are thermotropic, which does not allow us to extract the effect of purely steric forces. True hard body models that resemble top-shaped bodies are quite rare in the literature. Casey and Harrower[15] studied the stability of the smectic A and C phases in the system of coil-rod-coil triblock particles on a lattice using Monte Carlo simulation. They found that the smectic order can be stabilized by flexible terminal chains attached to the end of the mesogenic hard body. In a recent study, Varga and Fraden[16] studied the phase behavior of the triblock hard mesogens where the particle consists of three hard cylinder segments of different diameters. The main result of this work is that smectic C phase formation can be induced by increasing the diameter of the central mesogenic unit.

The goal of the present work is to study the effect of a central non-mesogenic unit in the formation of mesophases in a system of hard spherocylinders with an embedded hard sphere. As the spherical shape does not favor the liquid crystalline order, the central hard sphere will have a major influence on the phase behavior of the system. Our top-shaped hard body model makes it possible to determine explicitly the role played by the mesogenic (spherocylinder) part and the non-mesogenic unit (sphere) in the stability of different mesophases.

The paper is organized as follows. The molecular model is presented in Section 2; Section 3 is devoted to the Parsons-Lee theory of inhomogeneous hard body fluids. We show how to



implement the free energy calculations for the nematic, smectic and columnar phases in three subsections of Sec. 3. Technical details of the Monte Carlo simulation are given in Sec. 4. The phase diagram of the system of top-shaped molecules and the bulk properties of smectic and columnar phases are presented in Sec. 5. Finally, some conclusions are drawn in Sec. 6.

## 2.    Molecular model

We model the top-shaped molecule as a hard spherocylinder with an embedded hard sphere as shown in Fig. 1. The hard spherocylinder, which is a mesogenic unit, consists of a cylindrical core (length $L$ and diameter $D$) and two hemispheres with diameter $D$ enclosing the ends of the cylinder. The hard sphere with diameter ($\sigma$), which is a nonmesogenic unit, is placed at the center of the cylinder, which makes the top-shaped molecule symmetric. The diameter of the central unit is higher than the diameter of the cylinder ($\sigma > D$). With this particle shape we assume that only hard body interactions act between the particles. This means that all types of pairwise overlaps such as the rod-sphere, rod-rod and sphere-sphere are forbidden. In the description of the isotropic-nematic phase transition the top-shaped molecules are freely rotating, while for the treatment of the nematic-smectic, nematic-columnar and smectic A-smectic C phase transitions the molecules are assumed to be perfectly aligned in the direction of the nematic director. The examined mesophases are shown schematically in Fig. 2.

## 3.    Density functional theory

According to the Parsons-Lee extension of the second virial density functional theory (DFT), the free energy of an inhomogeneous hard body system can be written as a sum of ideal and excess contributions $\beta F = \beta F_{id} + \beta F_{ex}$, where

$$\beta F_{id} = \int d(\mathbf{1}) \rho(\mathbf{1}) \big[ \ln \rho(\mathbf{1}) - 1 \big], \tag{1}$$

$$\beta F_{ex} = \chi \int d(\mathbf{1}) \rho(\mathbf{1}) \int_{r_{12} \in OR} d(\mathbf{2}) \, \rho(\mathbf{2}). \tag{2}$$



In these equations $\beta = \dfrac{1}{k_B T}$ ($k_B$ being the Boltzmann constant and $T$ the temperature), $\rho(\mathbf{1})$ is the local number density, $(\mathbf{i}) = (\vec{r}_i, \vec{\omega}_i)$ is an abbreviation for the position and orientation of the particle $i$ and $\chi = \dfrac{4 - 3\eta}{8(1-\eta)^2}$ is the Carnahan-Starling prefactor. The packing fraction of the system is defined as $\eta = \rho v_0$, where $\rho = N/V$ is the number density and $v_0$ is the volume of the particle. In Eq. (2) $r_{12} \in OR$ means that the distance between two particles ($r_{12}$) must be in the overlap region as the particles are modeled as hard bodies. Using the symmetry properties of the examined mesophases, the free energy can be simplified substantially. In the following parts we present the suitable form of the excess free energy for nematic, smectic and columnar phases only, while the reformulation of the ideal free energy term for different mesophases is not discussed here as it is a simple problem.

### a) Isotropic and nematic phases

The main feature of the nematic phase is that the local density is position independent, i.e. $\rho(\mathbf{1}) = \rho(\vec{\omega})$ and the spatial integrals of Eq. (2) can be performed in advance. As a result, the excess free energy density becomes

$$\beta F_{ex} / V = \chi \int d\vec{\omega}_1 \rho(\vec{\omega}_1) \int d\vec{\omega}_2 \rho(\vec{\omega}_2) V_{exc}(\vec{\omega}_1 \vec{\omega}_2),\qquad(3)$$

where the excluded volume between two particles with orientation $\vec{\omega}_1$ and $\vec{\omega}_2$ is defined as $V_{exc}(\vec{\omega}_1 \vec{\omega}_2) = \int\limits_{r_{12} \in OR} d\vec{r}_{12}$. In the isotropic phase the above equation is even simpler as the local density is constant $\rho(\mathbf{1}) = \rho/4\pi$. As we are only interested in the determination of the stability limit of the isotropic with respect to the nematic phase, we simply perturb the isotropic density with the most dominant nematic function, which is the well-known second Legendre polynomial ($P_2$). Therefore the perturbed local density takes the form $\rho(\mathbf{1}) = \rho/4\pi\left(1 + \varepsilon P_2(\cos\theta)\right)$, where $\varepsilon$ is an infinitesimally small number and $\theta$ is a polar angle measured from the nematic director. Substitution of the perturbed local density into Eqs. (1) and (3) and expansion of the total free energy up to the first non-vanishing term in $\varepsilon$ gives the so-called bifurcation equation for the isotropic-nematic phase



transition[17]. Without going into the details we present the final equation for the packing fraction of the isotropic-nematic bifurcation, which is given by

$$\eta_{IN} = \frac{2 + x - \sqrt{(2+x)^2 - x(3+x)}}{3 + x}, \tag{4}$$

where $x = -20 v_0 / V_{exc,2}$ and $V_{exc,2} = 5 \int_0^{\pi/2} d\gamma \sin\gamma \, V_{exc}(\gamma) P_2(\cos\gamma)$. Note that Eq. (4) is valid for any hard body shapes, since the only input is the volume of the particle ($v_0$) and the excluded volume ($V_{exc}$). In some special cases, such the spherocylinder shape, the excluded volume is an analytical function of the angle between two particles ($\gamma$). However, for the top-shaped body we have been forced to determine it numerically due to the complicated form of the sphere-sphere, sphere-rod and rod-rod overlap regions.

**b) Smectic A and C phases**

A common feature of the untilted (A) and tilted (C) smectic phases is that the long ranged orientational order is accompanied by a one-dimensional positional order, i.e. the molecules are arranged in layers. The main difference between the tilted and untilted phases can be measured by the tilt angle ($\Psi$), which is the angle between the layer normal and the nematic director. For the smectic A phase the tilt angle is zero, while it is in the interval $0 < \Psi < 90°$ for the smectic C phase. In our study we use the perfect orientational order approximation for the description of positionally ordered phases. This means that the particles' long axes are always parallel and they all point in that special direction which forms an angle $\Psi$ with the $z$-axis of the Cartesian coordinate system. With this condition, the positional ordering takes place always in the direction of the $z$-axis, while the fluid structure survives in the $x$-$y$ plane for both smectic structures. The mathematical form of this approximation for the local number density can be written as $\rho(\vec{r}, \vec{\omega}) = \rho(z)\delta(\theta - \Psi)$. Exploiting the periodic property of the local density, i.e. $\rho(z) = \rho(z + d_0)$ where $d_0$ is the smectic period (or wavelength), the excess free energy density becomes



$$\beta F_{ex} / V = \frac{\chi}{d_0} \int_0^{d_0} dz_1 \rho(z_1) \int_{z_{12} \in OR} dz_2 \rho(z_2) A_{exc}(z_{12}, \Psi) \qquad (5)$$

where $A_{exc}(z, \Psi) = \int_{R_{12} \in OR} d\vec{R}_{12}$ is the excluded area of the overlap region at a given $z$ separation. Note

that the excluded area also depends on the tilt angle ($\Psi$), because the area of the slices cut out from the excluded volume is orientation dependent. However, the integral of the excluded area along the z-axis must be independent of the tilt angle as $V_{exc}^{\parallel} = \int_{z \in OR} dz A_{exc}(z, \Psi)$ is the excluded volume between two parallel molecules. To decide between the positionally disordered nematic phase and the smectic phases, one has to minimize the free energy with respect to the local number density and the tilt angle. To do this, we parameterize the density distribution with the following simple function[18]

$$\rho(z) = \rho d_0 \frac{\exp[\lambda \cos(2\pi z / d_0)]}{\int_0^{d_0} dz' \exp[\lambda \cos(2\pi z' / d_0)]} \ . \qquad (6)$$

where $\lambda$ measures the sharpness of the density peak. With this density ansatz the total free energy, which is now the sum of Eqs. (1) and (5), is a function of $\lambda$, $d_0$ and $\Psi$. Therefore the problem reduces to the minimization of the free energy with respect to these three parameters. Note that the ideal free energy term (Eq. (1)) favors the uniform density distribution, while the excess free energy (Eq. (5)) can reach very low values by proper packing of the particles along the $z$-axis. As a result, a subtle competition between the ideal and the excess free energies determines the stability of the smectic phases with respect to the nematic structure. The input of the minimization procedure is the excluded area, which is determined numerically at any tilt angle.

### c) Columnar phase

In the columnar phase the molecules are orientationally ordered and form a two dimensional lattice. The most efficient two dimensional packing can be achieved by a hexagonal structure, since the cross section of the molecule is a circle. Again we use the parallel approximation for the orientations of the particles' long axes, but we assume a hexagonal solid structure in the x-y plane, i.e. $\rho(\vec{r}, \vec{\omega}) = \rho(\vec{R})\delta(\theta)$, where $\vec{R}$ is a position vector in the x-y plane. Note that the particles are not



tilted in this phase. Using the fact that the system is uniform along the $z$-axis, it is easy to show that the excess free energy density is

$$\beta F_{ex} / V = \frac{\chi}{A} \int_A d\vec{R}_1 \rho(\vec{R}_1) \int_{R_{12} \in OR} d\vec{R}_2 \rho(\vec{R}_2) d_{exc}(R_{12}),$$ (7)

where $d_{exc}(R) = \int_{z \in OR} dz$ is the excluded distance between two particles and $A = \sqrt{3} d_0^2 / 2$ is the area of

the hexagonal unit cell. The relationship between the excluded distance and the parallel excluded

volume is $V_{exc}^{\parallel} = \int_{R \in OR} d\vec{R} \, d_{exc}(R)$, which shows that Eq. (7) gives the same nematic free energy in the

uniform limit ($\rho(\vec{R}) = \rho$) as the smectic free energy does in the same limit (Eq. (5)). The excluded

distance is an analytic function, but it has three different intervals due to the rod-rod, rod-sphere and

sphere-sphere exclusions. It is easy to show that

$$d_{exc}(R) = \begin{cases} 2L + 2\sqrt{D^2 - R^2}, & 0 \le R \le D \\ L + 2\sqrt{(\sigma + D)^2 / 4 - R^2}, & D < R \le (D + \sigma)/2 \\ 2\sqrt{\sigma^2 - R^2}, & (D + \sigma)/2 < R \le \sigma. \end{cases}$$ (8)

For the columnar order we resort to a two dimensional Fourier expansion of the density distribution

and we minimize the free energy with respect to the Fourier coefficients and the wave number (or

$d_0$). Considering the inversion symmetry of the columnar phase $(\rho(\vec{R}) = \rho(-\vec{R}))$, the density

distribution is represented by

$$\rho(\vec{R}) = \rho \sum_{\vec{k}} f_{\vec{k}} \cos[\vec{k}\vec{R}],$$ (9)

where $f_{\vec{k}}$ is the Fourier coefficient, and $\vec{k}$ is the wave vector. The details of the minimization

procedure can be found in our previous paper[19]. As in the case of smectic ordering, the competition

between the ideal gas contribution (Eq. (1)) and the excess free energy term (Eq. (7)) determines the

stability of the columnar phase with respect to the nematic structure. Furthermore, the stability of the

columnar phase with respect to smectic A and C phases can be determined from the comparison of the

free energy values at a given packing fraction, since the free energies of smectic and columnar

phases reduce to the nematic values in the uniform limit of the local density.



## 4.    Monte Carlo simulations

To test the theoretical predictions regarding the nematic-smectic transition we perform a series of constant pressure Monte Carlo (MC) simulations[20]. Model systems of $N$=256 parallel (i.e. "nematic") top-shaped particles are studied. The spherocylinder width is kept at $D = 1$ and the cylinder length is $L = 9$, so that the total particle length is $L+D = 10$ in all cases, while the central unit diameter is varied. The side ratio $c_z/c_x$ of the square prism ($c_x$=$c_y$) simulation cell is kept constant, with the z extension such that it will accommodate just four layers at the expected smectic density. The pressure is slowly incremented such that the packing fraction is increased from 0.1 to a value well beyond the respective phase transition density, i.e. $\eta \approx 0.7$ for the small diameter systems and $\eta \approx 0.5$ for $\sigma$=1.8-1.9 $D$. At each pressure $10^6$ MC cycles are performed, where each such cycle consists of $N$ trial particle moves and one attempted volume change.

The expected structural phase transitions are diagnosed by means of the dominant Fourier modes of the local particle density. Let $\rho(\vec{k}) = \dfrac{1}{V} \sum_{j=1}^{N} \exp\left[-i\vec{k}\vec{r}_j\right]$ with $V = c_x^2 c_z$ denote the Fourier component of the one particle density corresponding to a Fourier vector $\vec{k} = 2\pi\left(k_x / c_x, k_y / c_x, k_z / c_z\right)$ where $k_x$, $k_y$ and $k_z$ are integers. The periodic boundary conditions allow for certain discrete Fourier modes of the density. The zero vector is excluded, and for symmetry reasons one of the components, say $k_z$, may be restricted to positive values. For the same reason, if $k_z$=0, then we choose $k_y \geq 0$, and if $k_z$=$k_y$=0, then we let $k_x$>0. Considering these selection rules the total number of relevant Fourier vectors is $K = \left((2k_0 + 1)^3 - 1\right)/2$, where $k_0$ is an arbitrary upper limit for the absolute value of $k_x$, $k_y$, and $k_z$. We have chosen $k_0$=5, which yields a number $K$=665 of Fourier modes that are monitored at all times. Any inhomogeneity in the system will be indicated by enhanced values of structure factors pertaining to certain wave vectors. Writing $\rho(\vec{k}) = \rho'(\vec{k}) - i\rho''(\vec{k})$, the normalized structure factor $S(\vec{k}) = \left[\rho'(\vec{k})^2 + \rho''(\vec{k})^2\right]/\rho^2$ may vary



between 0 and 1. If $S(\vec{k})$ is small, the structure is more or less homogeneous along the respective $\vec{k}$. Smectic layering announces itself by a high value of some $S(\vec{k})$, where $\vec{k}$ is now the wave vector along the layer normal. If $\vec{k}$ points along the $z$ axis we have a smectic A, otherwise smectic C phase.

## 5. Results

First we present the results of the free energy minimizations and the phase coexistence calculations. In the case of the isotropic-nematic phase transition Eq. (4) yields the spinodal instability curve between the two phases. The free energies of the smectic and columnar ordering are obtained by minimization with respect to the variational parameters such as the period ($d_0$), tilt angle ($\Psi$) and Fourier amplitudes ($\lambda_0$, $f_k$). In the case of first order phase transitions, the coexisting densities of phases $\alpha_1$ and $\alpha_2$ are determined from the phase boundary conditions for the pressure ($P = -\dfrac{\partial F}{\partial V}$) and chemical potential ($\mu = \dfrac{\partial F}{\partial N}$), which are $P(\alpha_1)=P(\alpha_2)$ and $\mu(\alpha_1)=\mu(\alpha_2)$. On the other hand the transition density of second order phase transitions such as nematic-smectic A is given by the lowest density at which the smectic amplitudes vanish first, i.e. the inhomogeneous fluid becomes homogeneous. The results of the phase boundary calculations are collected in Fig. 3. Two molecular parameters govern the stability of the mesophases, one is the central sphere diameter ($\sigma$), while the other is the aspect ratio ($L/D$) of the mesogenic unit. As Figure 3 shows, $\sigma$ is responsible for the formation of columnar and tilted structures, while a sufficiently large shape anisotropy or aspect ratio ($L/D$) is very crucial in the stabilization of all mesophases. First of all we investigate the role of the aspect ratio in the phase diagrams. Comparison of the four panels of Fig. 3 demonstrates that the aspect ratio has to exceed five in order to stabilize the smectic and columnar phases. For example for $L/D$=9 and $\sigma/D$=1, which is the hard spherocylinder shape, the phase sequence is the usual isotropic, nematic, smectic and columnar. This is almost the same phase sequence as for freely rotating spherocylinders, since only the columnar phase is replaced by the solid phase due to the orientational fluctuations which are neglected in our calculations. Regardless of the value of the



central sphere diameter, increasing the aspect ratio stabilizes the nematic phase with respect to the isotropic one, as the particle becomes more anisotropic. In the limit of $L/D \to \infty$ the coexisting packing fractions of the isotropic-nematic phase transition go to zero[7], which creates very wide density regions for the mesophases. Interestingly the aspect ratio has a small effect on the stability regions of the observed smectic A, smectic C and columnar phases. It moves the regions of smectic A and smectic C phases in opposite directions by enhancing the stability region of the columnar ordering. The effect is more pronounced on the smectic C side, where the smectic C - columnar binodal moves in the direction of lower packing fractions. In the context of the above arguments we must mention that the isotropic-nematic phase transition is weakly first order, so the coexisting isotropic and nematic densities almost coincide with the presented spinodal curves.

Now we turn to the effect of the central sphere unit on the stability of mesophases. One can see that increasing $\sigma/D$ has a strong effect on all transitions irrespectively of the value of the aspect ratio. The nematic phase is destabilized with respect to the isotropic one with $\sigma/D$, because the molecule becomes more spherical and the packing entropy gain by orientational ordering is smaller. The same effect can be observed in the case of the smectic A phase. With increasing $\sigma$ there is less room for the particles in the layers to form a two-dimensional fluid and the packing of the mesogenic units is less efficient. This is due to the fact that the rod-sphere pairs do not match in the layer and the spheres give rise to extra unoccupied regions in the layers. As a result, the formation of the smectic A phase is shifted in the direction of higher density, and its stability range shrinks due to the formation of a columnar structure. The stabilization of the columnar order is also plausible, because the tendency of decreasing in-plane fluidity favors the formation of a two-dimensional solid structure in the layers. In addition, with increasing sphere diameter there is more room between the adjacent mesogenic units, which favors the fluidization of the system in the direction perpendicular to the solid layers. With increasing diameter the stabilization of the columnar phase is so strong that the smectic A phase completely disappears and a direct nematic-columnar phase transition takes place for $\sigma/D > 1.2$. However, further increase of $\sigma$ is not favorable for the columnar phase, because the



distance between the neighboring columns is of the order of $\sigma$, which also gives rise to large unoccupied regions in the space, i.e. the system cannot pack efficiently in the columnar phase with large central spheres. Therefore the system tends to be a new structure where it can preserve its fluidity to some degree and pack more efficiently. This structure is the smectic C phase, where the mesogenic units are tilted such that they almost lie inside the layer. With this arrangement the system becomes more packed as the mesogenic units get closer to each other and, in addition, the sytem becomes a two-dimensional fluid again. It can be seen in Fig. 3 that with further increase of $\sigma/D$ the tilted smectic structure becomes stable, while the region of the columnar phase moves to higher packing fractions. From these results we can conclude that the hard body shape has a very strong influence on the stability of different mesophases.

The stability of smectic ordering was also checked by NpT Monte Carlo simulation. By searching for the highest value of the structure factor $S_{max}$ and the corresponding wave vector it is possible to distinguish the nematic and smectic phases. In the simulation some criterion is needed to define the boundary of the smectic phases. We consider the phase as smectic A if the dominant density mode is along the $z$ axis, and the respective structure factor $S_{max}$ exceeds a value of 0.1. The latter condition may seem arbitrary, but experience shows that (a) the dominant wave vector, which is changing randomly at lower values of $S_{max}$, becomes robust and parallel to the $z$ axis as soon as $S_{max} > 0.1$; and (b) the rise of $S_{max}$ is so steep that the exact threshold value is of no importance. In contrast to smectic-A, the onset of the smectic-C phase was accompanied by a noticeable density jump at a given pressure. This is surprising, since the transition is also expected to be of second order. A possible explanation is the discrete distribution of allowed Fourier modes in the finite, periodic simulation cell. It appears that within a few compression steps the tilted layers are "locking in" with one of these modes, resulting in a simultaneous rise of the packing fraction and the structure factor. We use the values of $\eta$ immediately before the steep rise to define the nematic-smectic C transition density. Fig. 4 shows the density dependence of the structure factor for $\sigma/D=1.0$, 1.02, 1.04 and 1.06 (a) and for $\sigma/D=1.80$, 1.85 and 1.90 (b) (in all cases we set L/D=9). In panel (a) we can see



that the smectic phase has lower and upper bounds in all cases. The wave vector ($\vec{k}$) points along the z axis, and therefore the lower bound is due to the second order nematic-smectic A phase transition as $S_{max}$ increases smoothly from zero, while the upper bound is a first order phase transition between the smectic A and the columnar phases. Note that $S_{max}$ jumps suddenly and there is also a jump in the equation of state at the upper bound. This is an indication of the first order character of the smectic-columnar phase transition. It can be seen in Fig. 4 (a) that the density range of $S_{max}>0.1$ shrinks with increasing diameter of the central core, i.e. the smectic A phase becomes less stable as $\sigma$ gets larger. The results of the Monte Carlo simulations for small diameters of the central unit are collected together with the corresponding DFT results in Table 1, where we can see that the smectic phase can be destabilized with respect to nematic and columnar phases by increasing the diameter of the central core. Comparison of the simulation and theoretical results shows good agreement. In addition, the failure of our mean field density functional theory can also be seen in Table 1, as the upper and lower borders of the smectic A phase are underestimated and the smectic period is too large. This is due to the misrepresentation of the density correlations in our theory. The recent study of Capitán et al.[21] shows that the inclusion of the density correlations into the theory does help to get a better equation of state and smectic period, but it does not improve the phase diagram. This was demonstrated in the system of parallel hard cylinders, which is very close to our systems with $\sigma/D$=1. As mentioned before, we have also performed Monte Carlo simulations in a system of particles with $L/D$=9 and $\sigma/D\geq1.8$ to check the prediction of our DFT study for the stability region of the smectic C phase. Panel (b) of Fig. 4. shows that $S_{max}$ increases steeply as the density is increased beyond the transition point and then levels off. The wave vector is now pointing along directions different from the z axis (see Table 2), as corresponds to the formation of a tilted smectic phase instead of an untilted one. The nematic-smectic C transition properties of the MC simulations and DFT calculations are compared in Table 2. The agreement is surprisingly good for the packing fractions and tilt angles, which may be due to the fact that the dominant interaction in the tilted layers is the hard sphere-hard sphere exclusion, which is accurately represented by the Carnahan-Starling prefactor in the theory.



However, the lack of proper inclusion of the density correlations is still present, because the smectic period is again overestimated. Another important difference is that we have not found an upper limit for the stability of the smectic C phase using MC simulations, i.e. no smectic C-columnar phase transition is observed by increasing the density. If present, the columnar phase appears at higher densities than those tested in our simulation. It is also plausible that the columnar phase is not stable for these values of $\sigma$. The phase diagrams depicted in Fig. 3 show that at a given aspect ratio there is probably a maximum $\sigma$ for which a columnar phase can be found (a possible vertical asymptote).

Now we turn to the discussion of the structural properties of the smectic and columnar phases along the transition curves, and we also examine the density dependence of these properties. The smectic (columnar) period is an important parameter because it reflects the average distance between the neighboring layers (columns). The values of $d_0$ at the nematic-smectic and nematic-columnar phase transitions are shown in Fig. 5 for a system of particles with $L/D=9$ and various $\sigma$. It can be seen that the smectic A phase has the usual layered structure with a period that is proportional to the length of the mesogenic unit ($L$). The effect of the increasing diameter ($\sigma$) is to decrease the smectic period, i.e. the neighboring layers get closer to each other. This is quite reasonable because the average distance between the mesogenic units inside the layers increases with $\sigma$, i.e. the system can pack more efficiently if the particles of the neighboring layers are closer to each other. Interestingly we have not observed strong interdigitations between the layers. In the case of 11-site Lennard-Jones (LJ) top-shaped rigid particles, Kim et al.[9] have observed, for $\sigma/D>1.7$, smectic periods of half the mesogenic length. The reason why they could observe such strong interdigitation of the layers may be attributed to the fact that 3 large LJ sites in the middle of the linear array strongly favor a side-by-side parallel arrangement instead of the slipped parallel one, i.e. the model of Kim et al. prefers the smectic A phase to smectic C even for very large $\sigma$.

The next phase along the phase coexistence curve is the columnar structure (Fig. 5). We can see that the columnar period is now in the order of $\sigma$, which is not surprising as the interactions between the columns are dominated by the hard body exclusions between the central spherical units.



This argument is supported by the fact that the columnar period is a more or less linearly increasing function of $\sigma$. At higher $\sigma$ the columnar phase is replaced by a tilted smectic ordering. In this phase the smectic period is very low, but it is always higher than the columnar period. Interestingly it depends only weakly on the diameter of the central unit but strongly on the length of the mesogenic unit. This effect can be understood by examining the aspect ratio dependence of the tilt angle along the nematic-smectic C phase transition (See Fig. 6). In all studied cases the tilt angle increases first with $\sigma$ and then starts to decrease. It can be seen also that the aspect ratio has a strong influence on the tilt angle and consequently on the smectic period. With increasing the length of the mesogenic unit the particles are more tilted with respect to the layer normal. This effect may be induced by the hard bodies staying in the interstitial regions as the interstitial particles exclude a smaller volume in the neighboring layers if the particles are more tilted. This is especially true for longer mesogenic units. Therefore the molecular parameter dependence of the tilted smectic structure (tilt angle and smectic period) is a complicated interplay between the in-layer packing and the interlayer packing tendencies. The packing fraction dependence of the smectic and columnar periods is more understandable (see Fig. 7-8). The period of both smectic A and columnar phases decreases with increasing density. This is simply a consequence of the fact that the only way to pack more particles into very dense layers (columns) is to decrease the distance between the neighboring layers (columns). In the case of a tilted smectic phase, not only the compression of the layer thickness helps to increase the packing fraction, because the increasing tilt angle can also give rise to extra free room for the particles to accommodate in. Therefore, the smectic period is a decreasing function of the packing fraction, while the tilt angle increases with it (Fig. 8). Finally, in Fig. 9 we show the density profiles of the positionally ordered phases at the same packing fractions and aspect ratio of the mesogenic unit ($L/D$=9), but at different sphere diameters. Comparison of the smectic A and C density profiles shows not only the differing smectic periods, but also the difference in the sharpness of the peaks. While the smectic A ordering is very peaked, the tilted smectic structure is accompanied by a smooth density distribution with nonzero interstitial density. This also proves that



the structure of the smectic C phase (tilt angle and the smectic period) is strongly affected by the interstitial particle-mediated interactions between neighboring layers. In addition, it can be seen that the structure of the columnar phase is hexagonal; the density profile is very peaked and very few particles stay between the columns.

## 6.    Conclusions

In this paper we have studied the effect of varying the central hard sphere diameter on the mesophase forming ability of uniaxial hard rod-fluids. Our theoretical calculations and simulation study show that the non-mesogenic central unit destabilizes both the nematic and smectic A phases. With increasing sphere diameter the top-shaped particle becomes more spherical, and the packing entropy gain arising from orientational and positional ordering becomes less with respect to the loss of orientational entropy, i.e. the nematic and the smectic ordering takes place at higher packing fractions. However, the spherical central core gives rise to the formation of columnar and smectic C ordering. The columnar phase is very stable in the interval of $1.2 < \sigma/D < 1.7$ with a first order phase transition from the nematic phase, while the smectic C phase becomes stable for $\sigma/D > 1.7$ for most of the studied aspect ratios. The shape and size incompatibility between the spherical and rodlike units is the source of the detected phase behavior. The rodlike units attempt to form layers, but the spheres make the layers too spacious. Consequently the columnar phase becomes more packed and stable than smectic A. However, the packing efficiency of columnar ordering depends very sensitively on the diameter of the central unit, i.e. the columnar order becomes less packed for larger diameters, and the system undergoes a phase transition from nematic to smectic C. We expect these mesophases to be absolutely stable, although we should bear in mind that solid phases may appear in the system at high enough densities, and could partially modify the phase diagrams reported in this paper.

Our results show that hard body models of mesogenic systems are able to account for the effect of increasing central core on liquid crystalline order. In agreement with our findings, the replacement of hydrogen atoms in the central core of mesogenic molecules with bulkier groups (e.g.



methyl, ethyl, NO₂) generally results in destabilization of nematic and smectic A phases[22]. In addition, it is shown in the recent experimental study of Tian et al.[23] that the stability of mesophases depends very sensitively on the central core. The phase behavior of three-segment coil-rod-coil molecules shows dramatic changes when lateral methyl or ethyl groups are added to the centre of the mesogenic molecule. In the case of methyl groups hexagonally perforated layers are formed, while the ethyl group gives rise to columnar order[23]. In some cases tilted layered structures are also found. Even though our molecular model is simple, it produces quite reasonable explanations for the observed phase behavior of bulky mesogens.



# Acknowledgements

We are grateful to Giorgio Cinacchi for useful discussions and for bringing to our attention the work of Geoffrey R. Luckhurst. We would like to thank E. Velasco and L. Mederos for their practical help and advice to this paper. SV and FV would like to thank the financial support of the Austrian-Hungarian Action Foundation (grant number: 77öu11). DH acknowledges the support of the Spanish Ministry of Education through contract No. EX2009-0121.

# Figures

**Figure 1)**

Hard body representation of a symmetric top-shaped molecule, consisting of a hard spherocylinder of cylinder length $L$ and diameter $D$ and an embedded hard sphere of diameter $\sigma$ placed in the center of the mesogenic unit.

**Figure 2)**

Schematic representation of isotropic (a), nematic (b), smectic A (c), smectic C (d) and columnar (front view: e, top view: f) phases.

**Figure 3)**

Effect of varying the hard sphere diameter ($\sigma$) on the phase transition of top-shaped molecules in the $\eta$-$\sigma/D$ plane for aspect ratios ($L/D$) of 5 (a), 9 (b), 13 (c) and 17 (d). The dashed curve represents the continuous nematic-smectic phase transitions, the dot-dashed curve is the border between the isotropic and nematic phases as computed by bifurcation analysis, and the solid curves show the binodals between nematic, smectic and columnar phases. The shaded areas are the two-phase regions. The squares indicate the position of the critical end points. The isotropic, nematic, smectic A, smectic C and columnar phases are denoted as $I$, $N$, $Sm_A$, $Sm_C$ and $C$, respectively.

**Figure 4)**

The largest Fourier mode of the density correlation ($S_{max}$) as a function of packing fraction for $\sigma/D$=1, 1.02, 1.04, and 1.06 (a) and $\sigma/D$=1.80, 1.85 and 1.90 (b). The curves are the results of NpT Monte Carlo simulation. The aspect ratio ($L/D$) is 9.

**Figure 5)**

Smectic or columnar period ($d_0/D$) vs. the diameter of the central hard-sphere ($\sigma/D$) along the transition curve between nematic and positionally ordered phases in the system with $L/D = 9$

**Figure 6)**

Tilt angle of the smectic C phase ($\Psi$) vs. diameter of the central hard-sphere ($\sigma/D$) along the nematic-smectic C phase transition curve for different aspect ratios ($L/D$).



**Figure 7)**

Packing fraction dependence of smectic A and columnar periods ($d_0/D$) in the system with $L/D = 9$ and $\sigma/D = 1.1$. Note the broken scale on the vertical axis.

**Figure 8)**

Packing fraction dependence of the period ($d_0/D$) and the tilt angle ($\Psi$) in the system with $L/D = 9$ and $\sigma/D = 1.9$ are shown in (a) and (b), respectively.

**Figure 9)**

Local packing fraction of the smectic A (a), columnar (b) and smectic C (c) phases for the aspect ratio $L/D=9$ and packing fraction $\eta=0.45$. The diameter of the central hard sphere ($\sigma/D$) is 1.1 in panel (a), 1.5 in panel (b) and 1.9 in panel (c).

**Table 1)**

Diameter ($\sigma/D$) dependence of the stability range of the smectic A phase. The data are the results of NpT Monte Carlo simulation (MC) and those of density functional theory (DFT). $\eta$ is the packing fraction (we include the range where smectic ordering may be observed) and $d_0/D$ is the smectic period. The aspect ratio is $L/D=9$.

**Table 2)**

Summary of the Monte Carlo simulations and the density functional calculations of a system of top-shaped particles with $L/D=9$. The diameter of the central unit ($\sigma/D$), the packing fraction at the nematic-smectic C phase transition ($\eta$), the integer components of the wave vector ($k_x,k_y,k_z$), the tilt angle ($\Psi$) and the smectic period at the nematic-smectic C phase transition ($d_0/D$).



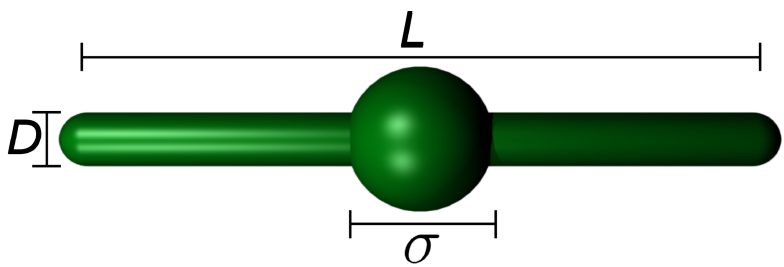

**Figure 1.**

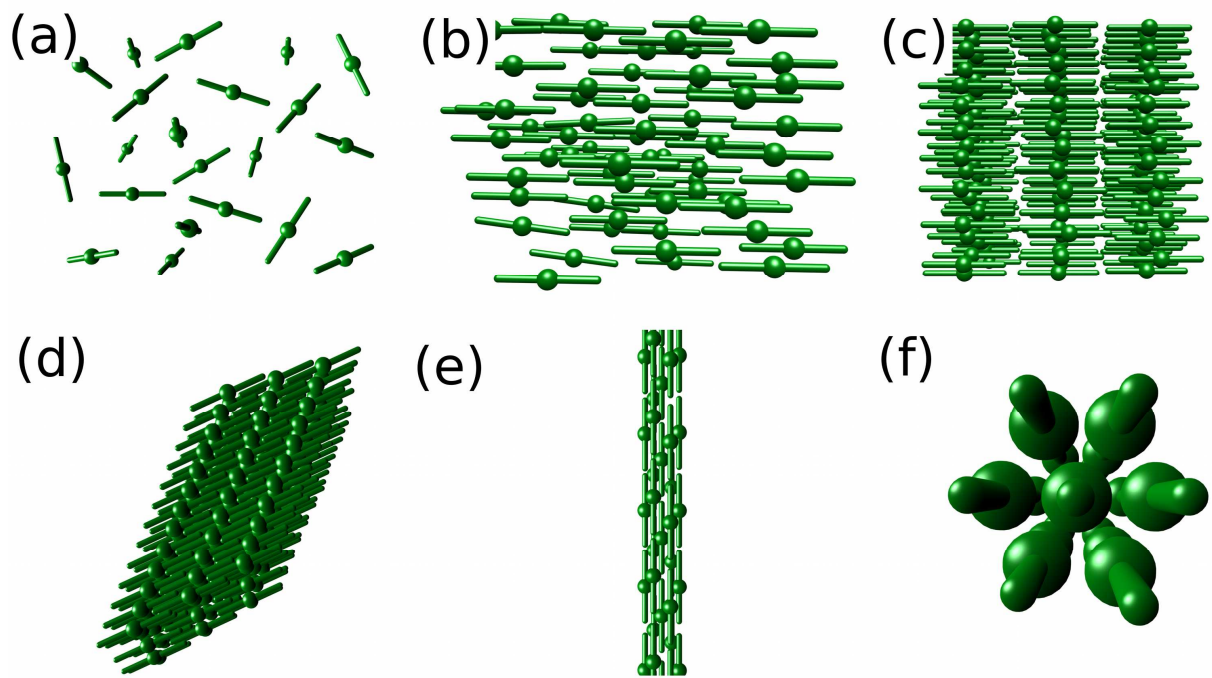

**Figure 2.**



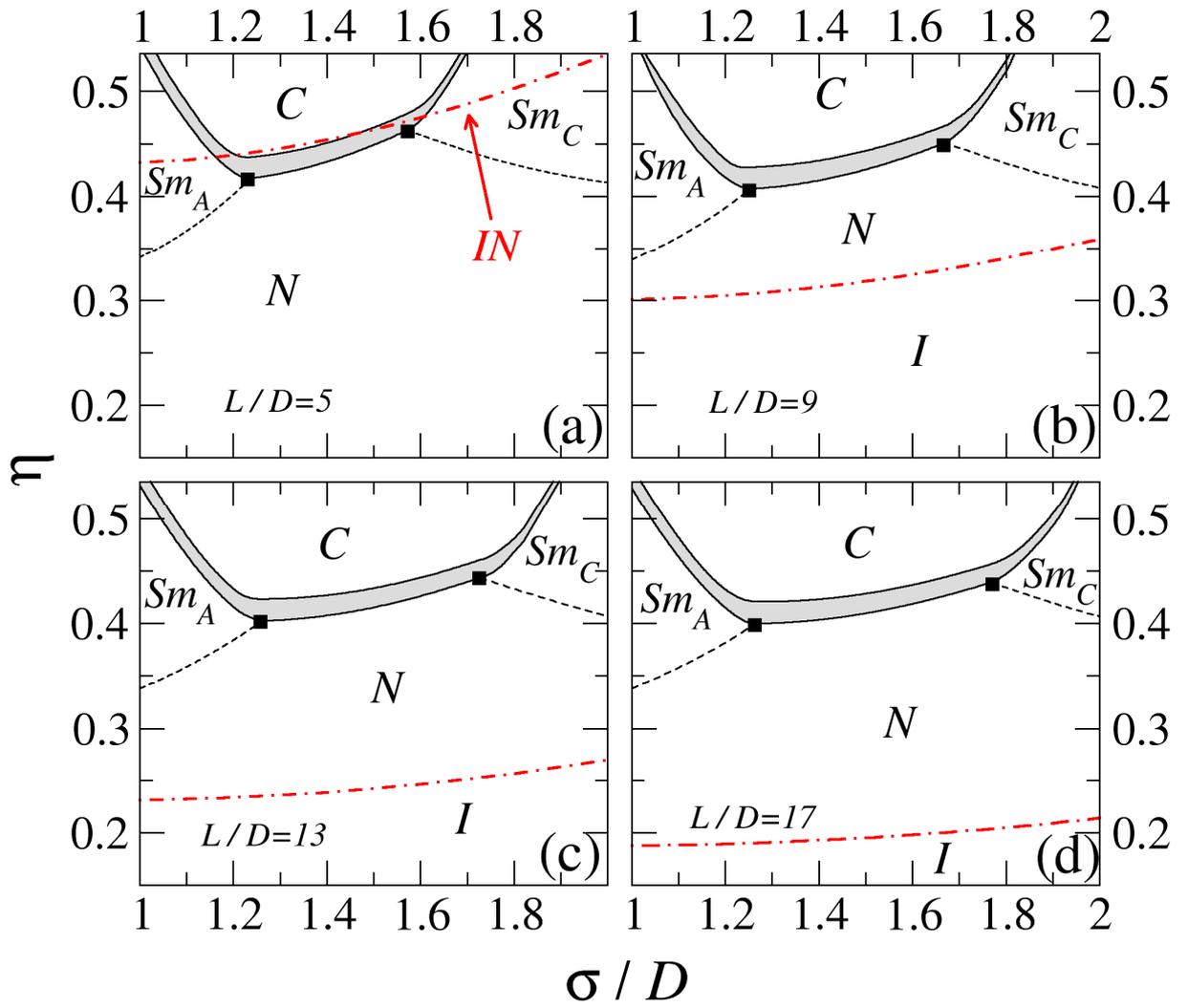



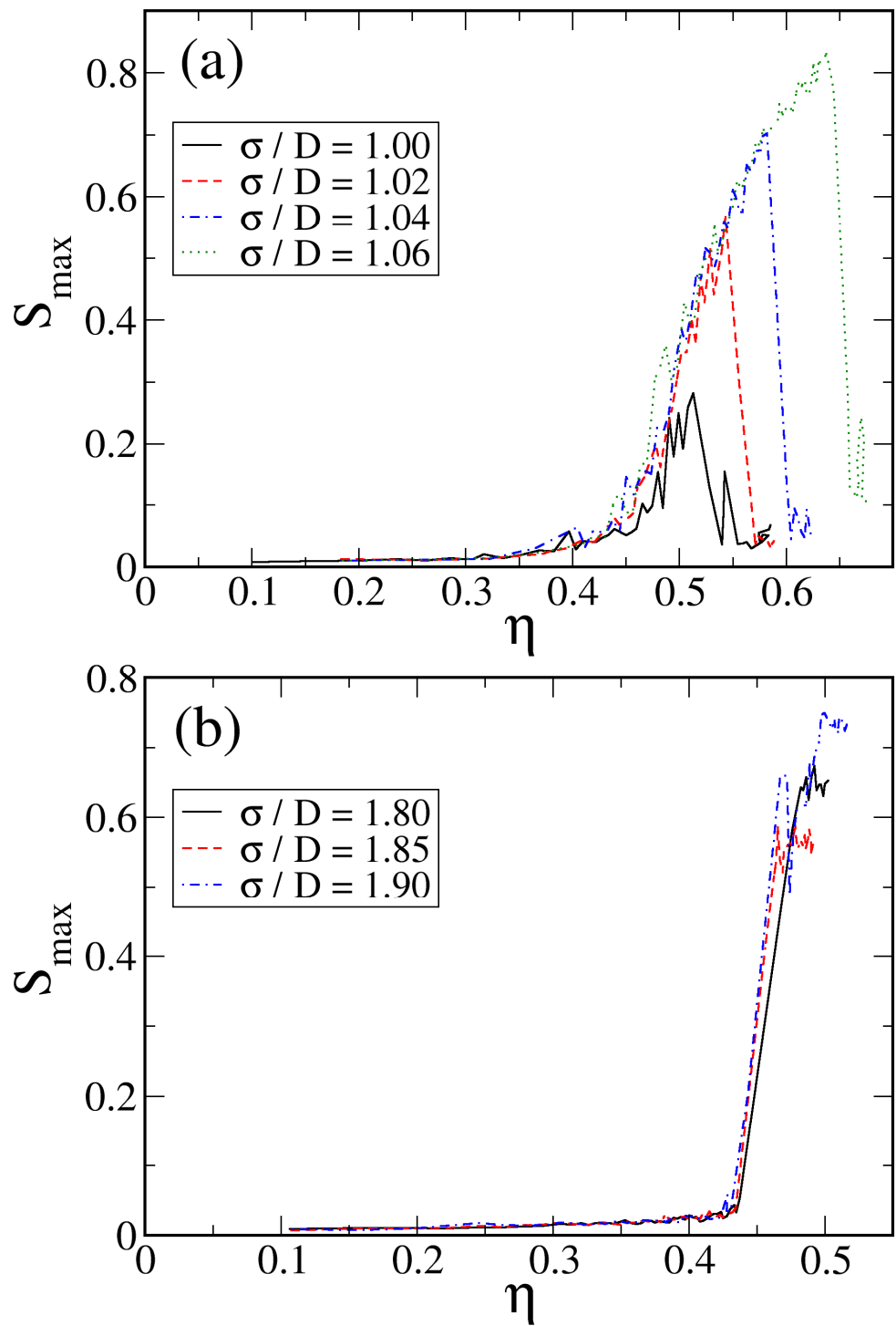

**Figure 4.**



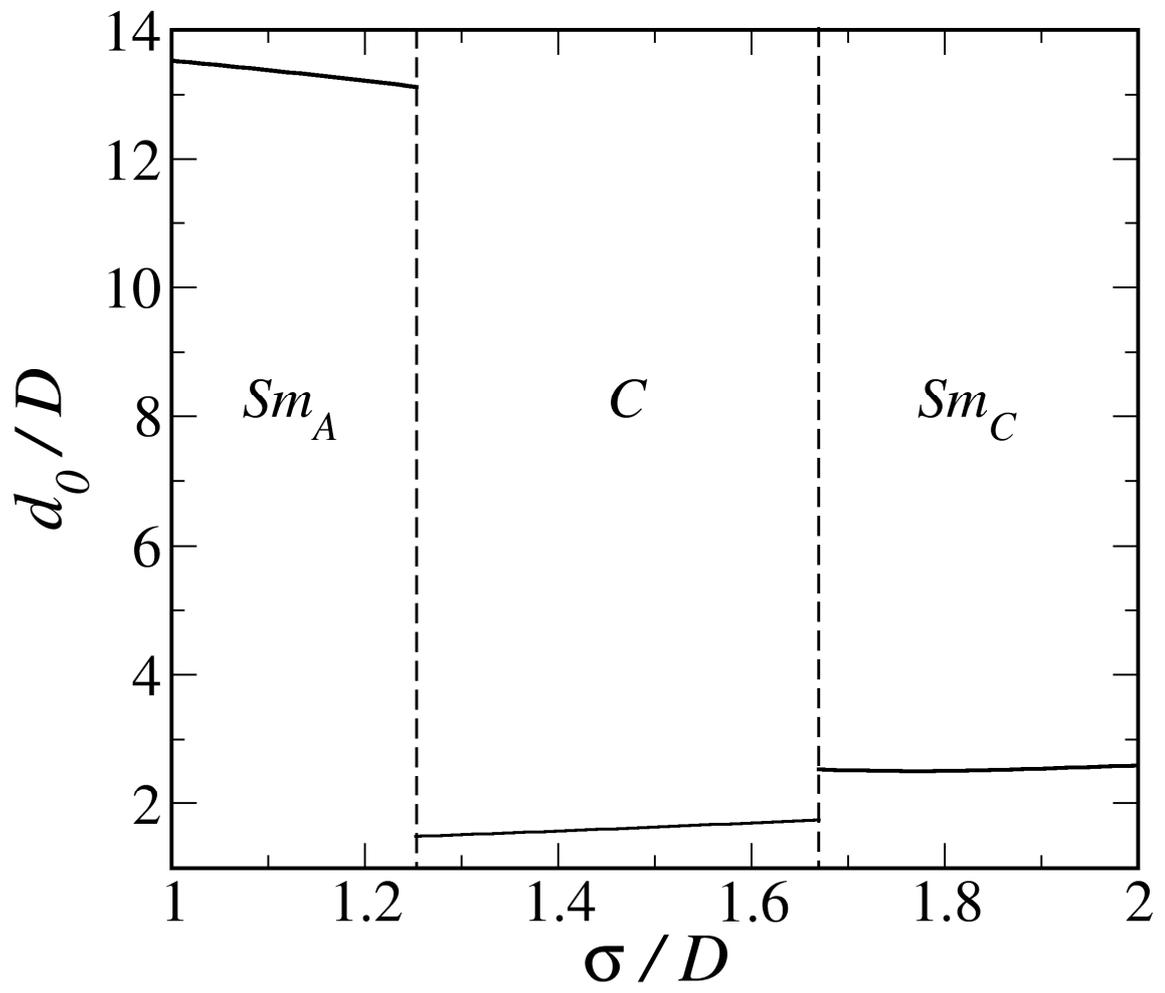

**Figure 5.**



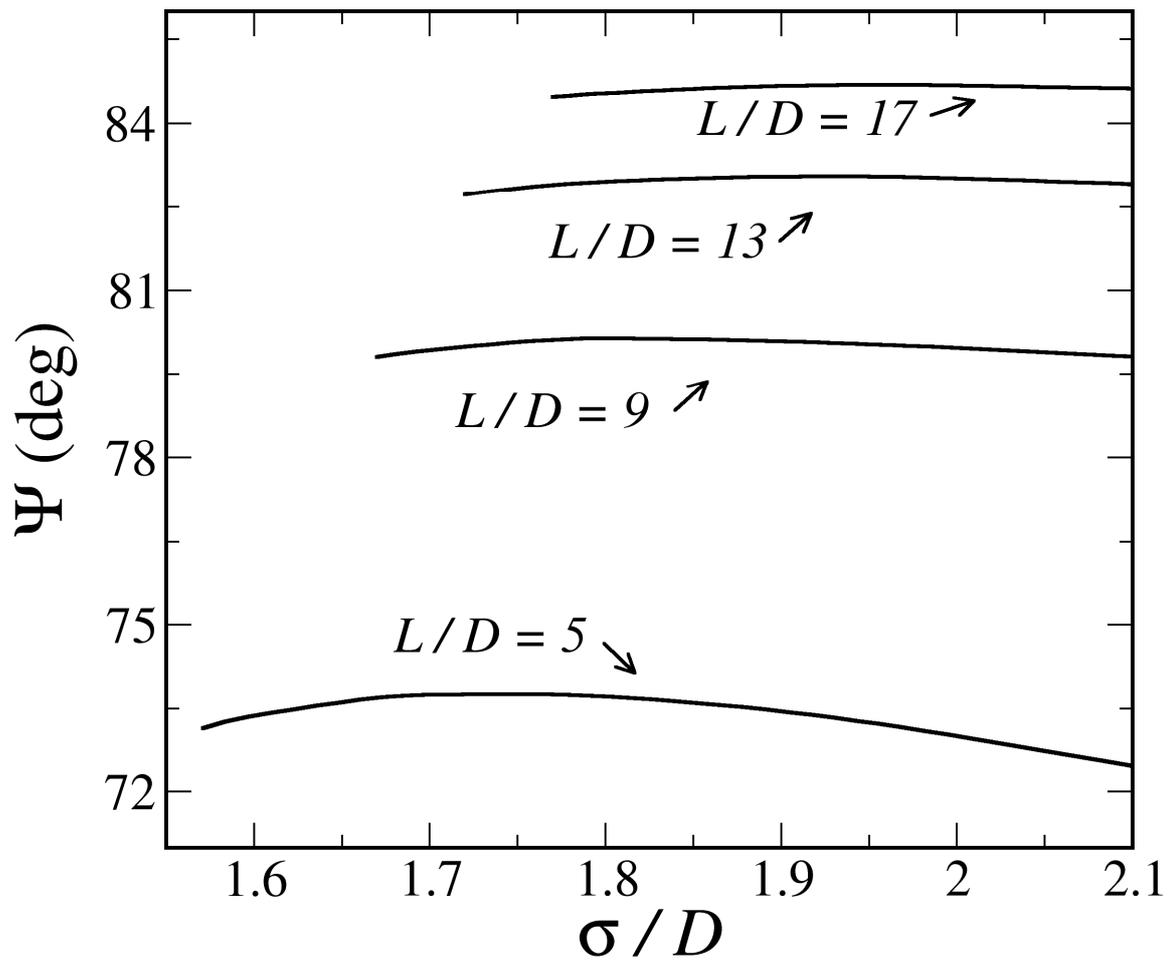

**Figure 6.**



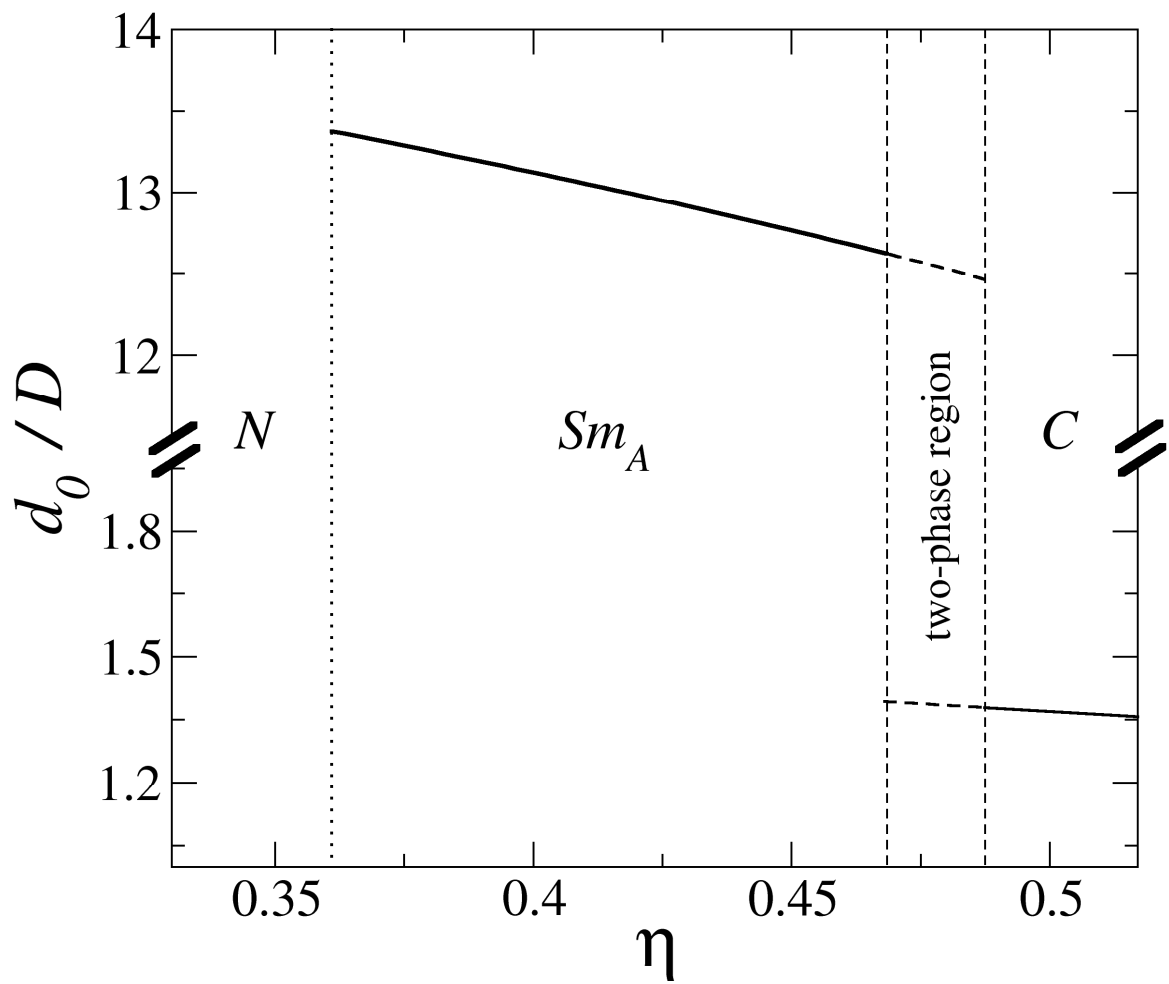

**Figure 7.**



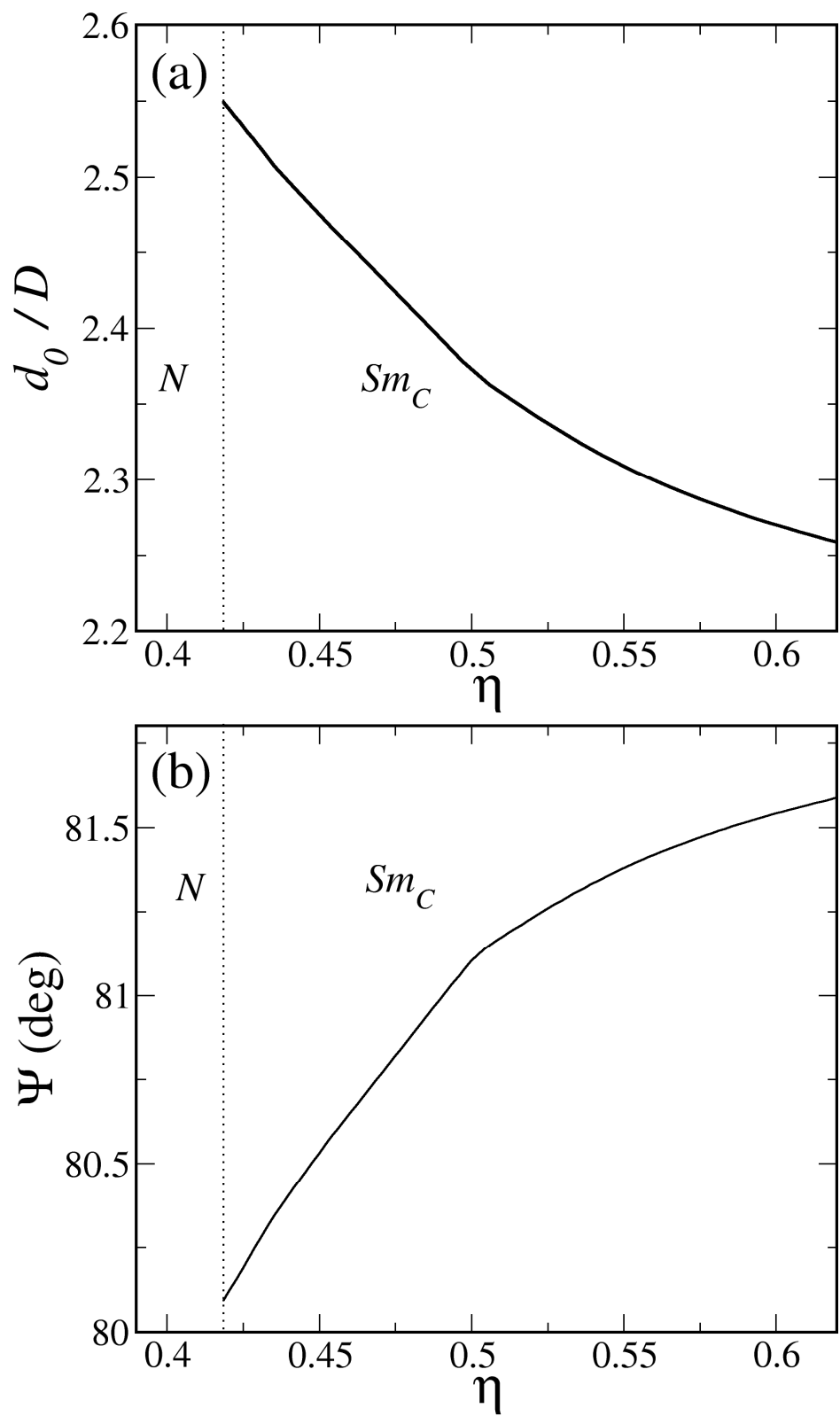

**Figure 8.**



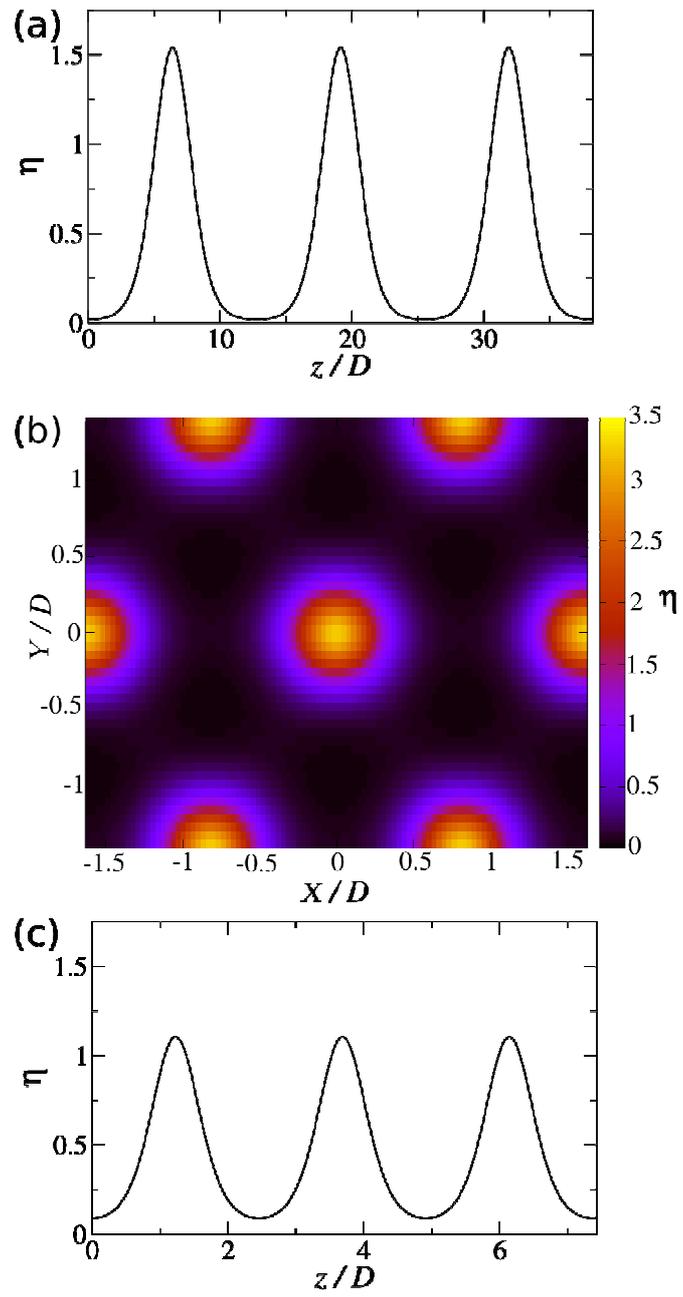

**Figure 9.**



| $\sigma/D$ | MC | | DFT | |
|---|---|---|---|---|
| | $\eta$ | $d_0/D$ | $\eta$ | $d_0/D$ |
| 1.00 | $0.44 - 0.66$ | $11.5 - 9.9$ | 0.34-0.56 | 13.52-12.13 |
| 1.02 | $0.45 - 0.60$ | $11.2 - 10.2$ | 0.34-0.54 | 13.50-12.24 |
| 1.04 | $0.46 - 0.57$ | $11.0 - 10.3$ | 0.35-0.51 | 13.47-12.33 |
| 1.06 | $0.47 - 0.53$ | $10.9 - 10.5$ | 0.35-0.50 | 13.44-12.42 |

**Table 1.**

| $\sigma/D$ | MC | | | | DFT | | |
|---|---|---|---|---|---|---|---|
| | $\eta$ | $k_x/k_y/k_z$ | $\Psi$(deg) | $d_0/D$ | $\eta$ | $\Psi$(deg) | $d_0/D$ |
| 1.80 | 0.44 | -5/2/4 | 80.6 | 1.9 | 0.43 | 80.14 | 2.51 |
| 1.85 | 0.43 | -3/5/4 | 81.3 | 1.8 | 0.42 | 80.12 | 2.53 |
| 1.90 | 0.43 | -5/2/4 | 80.6 | 1.9 | 0.42 | 80.09 | 2.55 |

**Table 2.**